# A Trust Management Framework for Vehicular Ad Hoc Networks

Rezvi Shahariar and Chris Phillips

School of Electronic Engineering and Computer Science, Queen Mary, University of London, London, UK

## Abstract

*Vehicular Ad Hoc Networks (VANETs) enable road users and public infrastructure to share information that improves the operation of roads and driver experience. However, these are vulnerable to poorly behaved authorized users. Trust management is used to address attacks from authorized users in accordance with their trust score. By removing the dissemination of trust metrics in the validation process, communication overhead and response time are lowered. In this paper, we propose a new Tamper-Proof Device (TPD) based trust management framework for controlling trust at the sender side vehicle that regulates driver behaviour. Moreover, the dissemination of feedback is only required when there is conflicting information in the VANET. If a conflict arises, the Road-Side Unit (RSU) decides, using the weighted voting system, whether the originator is to be believed, or not. The framework is evaluated against a centralized reputation approach and the results demonstrate that it outperforms the latter.*

## 1. Introduction

Vehicular Ad Hoc Networks (VANETs) can provide traffic advice, safety announcements, and infotainment services to road users. Typically, a vehicle may report an emergency event with other road users or may request the location of a petrol pump or a nearby parking area. VANETs are also deployed to mitigate the aftereffect of road incidents and to warn vehicles in advance. However, as this application involves wireless communication, it is at risk of security attacks. Additionally, drivers can fraudulently broadcast false messages. To be successful, messages must be accurate and trustworthy, otherwise, with a malicious untrue message, a vehicle can mislead many others causing congestion or other undesirable phenomena.

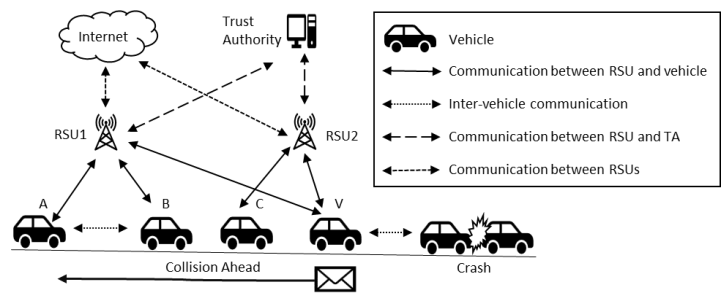

Figure 1. Typical Mechanism for Reporting an Accident

For example, as shown in Figure 1, suppose a vehicle "V" broadcasts a message reporting a crash. This message could be truthful or false. If other vehicles receive a false message, their subsequent detour will impact their travel time. However, for a truthful announcement, the detour permits them to avoid potential congestion.

In a VANET, outsider attacks can be thwarted using a cryptographic scheme but not insider attacks. Trust-based approaches are used to thwart insider attacks from malicious authorized users [1, 2, 3]. It is noted in [4, 5] that trust schemes can improve security by identifying dishonest vehicles and revoking messages from them. Even so, trust approaches cannot protect VANETs completely [1]. Basically, the trust that vehicle W attributes to vehicle V is the confidence W places in a set of actions from V. Typically, the reliability of relayed information is periodically evaluated using predefined metrics and computational methods [6]. Vehicles that consistently maintain a good trust score can be considered trustworthy by others as their current trust scores rely on their previous trustworthy announcements. However, it is not guaranteed that a trusted vehicle will always broadcast trustworthy messages.

In existing approaches [2, 3, 6-8] both trusted and untrusted vehicles can broadcast messages. Untrusted vehicles are expected to broadcast more malicious messages than trusted vehicles, which produces an additional demand on the network both in terms of message volume and the verification process. This places a considerable burden on the receivers. Methodologies based on direct and indirect trust require regular monitoring of activities across both single and multi-hop transmission ranges. Some approaches [2, 4, 9] can result in considerable trust metric exchanges to verify the original announcement. These messages, along with the event announcement, complicate the situation as it is necessary to evaluate their validity in a short time frame due to fast vehicle movement [10]. The authors in [6] claim receivers should decide the trust of messages in a short timeframe. However, when receivers independently compute trust from their neighbours' trust metrics, they suffer from high response times [2, 6, 9]. Alternatively, approaches that allow trust computation at a centralized server need to communicate to obtain updated trust information concerning the sender vehicle. This introduces an additional delay in the decision-making process concerning emergency events. Consequently, some vehicles may drive into the event zone despite being previously warned, as suggested in [11]. Also, there is an open debate [4, 8] regarding how often a centralized server should communicate revised trust data. Therefore, this paper proposes a novel Tamper-Proof Device (TPD)-based sender-side trust management framework for VANETs with the following features:

- To the best of our knowledge, for the first time, this framework employs a sender-side trust model to regulate access control using information accuracy, delay, and positional differences collected from the sender vehicle itself. Unlike other approaches [2, 4, 7, 8, 10, 12, 13], there is no flow of trust metrics unless a reporter vehicle refutes an announcement.
- Various classes of messages along with their associated trust threshold are defined for regulating access control which confirms that only trusted vehicles can announce messages.
- The scheme employs a collaborative untrue message discovery algorithm for detecting various forms of attack.
- Receivers can instantly act on a sender's message knowing that the sender must have sufficient trust, or they can report an untrue attack if they do not see the event on the roads.
- The framework is simulated in the Veins to verify its satisfactory operation. The communication overhead and response time are compared against a typical reputation approach [8] with varying vehicular densities and speeds. Moreover, the accuracy of the framework is presented with differing percentages of malicious and benevolent feedback.

Our paper is organized as follows. Section 2 reviews existing trust models. Section 3 presents the core parts of the proposed trust model. Section 4 illustrates the system model and verification of the framework. Section 5 then analyses the framework and provides a performance comparison against a baseline approach. Finally, in Section 6, we conclude this work.

## 2. RELATED WORK

Till now, many trust models for VANETs have been proposed for evaluating message trustworthiness [5]. These models are categorized into three groups. The first group is called the Entity-Oriented Trust Model (EOTM) which verifies only an entity's trust. The second group is called the Data-Oriented Trust Models (DOTM) which focus on evaluating the trustworthiness of data only. The third group is called the Hybrid Trust Model (HTM) which evaluates both entity's trust and the reliability of data. Below trust models from these categories are highlighted briefly.

Entity-oriented trust models are typified by [7], where the researchers use a Tamper Proof Module (TPM) on every vehicle to find the cost for the transmission and then adjust trust from the receivers' feedback. However, the trust score updating leads to excessive communication. Conversely, the authors in [14] apply fuzzy logic to calculate trust using experience, plausibility, and location accuracy. It can detect bogus messages and alteration attacks. However, additional communication is required as vehicles consult with fog nodes to check the location accuracy. Ref [9] uses both fuzzy logic and Q-learning for trust calculation. This approach is evaluated in terms of precision and recall with varying numbers of malicious vehicles. However, the model requires repeated sensing of messages from neighbours. In contrast, Ref [2] uses the Bayesian rule and Dempster-Shafer Theorem (DST) for trust calculation. It combines independent beliefs to determine the trust of a vehicle. However, an erroneous recommendation can bias the trust calculation. Reference [4] uses reputation and receivers' feedback on received messages to calculate trust. The scheme is evaluated in presence of false messages in both urban and highway scenarios. Nevertheless, it may suffer from excessive trust metric dissemination. The researchers in [15] propose a past interactions-based reputation management scheme for VANETs. A vehicle collects a signed reputation from the server to attach to its messages. A receiver verifies the message to accept or reject it and updates the reputation of vehicles on the server. However, this approach requires periodic reputation exchanges with the server. Another trust model is proposed in [16]. They consider familiarity, packet delivery ratio, timeliness, and interaction frequency as parameters to manipulate final trust. This model is analyzed considering the recent history of interactions but not considering any attacker model. In [17], the authors introduce a blockchain-based trust management protocol for VANETs to update of trust of vehicles as low, medium, and high at the Trust Authority (TA) using tamper-proof logs periodically. However, this is not validated using any known attack.

Data-oriented trust models are typified by [18], where packets are forwarded along the most trusted path using a trusted routing protocol. An intrusion detection module thwarts only denial of service (DoS) attacks. Ref [19] presents an infrastructure-less, data-oriented trust model which verifies content similarity and conflict as well as route similarity. Conversely, the reference [20] proposes a distance and geolocation-based probabilistic approach to estimate the trust from received data. This model does not forward a message beyond a certain distance. Ref [21] proposes a Bayesian Inference-based voting mechanism and vehicles run Dijkstra's algorithm as a route update requires upon every message arrival. Messages carry a time parameter as road ID while being forwarded. However, this requires frequent maintenance of routes. The researchers in [22] propose a trust model called FACT for achieving reliable information dissemination in VANETs using two modules. The first module evaluates the trust of the message, whereas the second one finds a highly trusted path for forwarding messages. However, this is an application-oriented scheme that does not use any infrastructure to monitor activities.

The researchers in [5] evaluated one data-oriented, one entity-oriented, and one hybrid trust model under various adversary scenarios. In addition, a risk assessment model is also presented for the identification of critical vulnerabilities. Ref [6] checks the reliability of messages using direct interactions and stores previous interactions and the trust of neighbours in a local database. However, there is no false trust message detection scheme. Conversely, the authors in [12] calculate data trust from multiple vehicle responses and vehicle trust from functional and recommendation trust. Though this model considers simple, bad mouth and zigzag attacks, but not compared with any model. In [3], RSUs use hash message authentication code (HMAC) and digital signature to evaluate the trust of vehicles based on neighbour trust values and only reward. They measure communication overhead and suggest integrating the ID-based and batch signatures in the future. The researchers in [8] update vehicle reputations at the RSUs using the noticed events from vehicles. The RSU next announces updated reputations to vehicles. Receivers store all the messages about an event until a timer expires to decide whether an event is true or false. However, this model suffers from high response times and communication overhead. Alternatively, in [23], a self-organizing trust model is considered for both urban and rural settings which can detect fake locations, and fake times and revoke messages as necessary. This model validates the trust of messages and then accepts the message with the highest trust for an event. However, its efficacy is not analyzed. In [24], the authors present a trust model, where the entity-centric trust model of this scheme thwart the black-hole attack and selective forwarding attack. The data-centric trust model is used to discover relations among data and performs trust evaluation based on utility theory. The data trust model can be further improved by selecting the appropriate utility parameter. Alternatively, in [25], a risk-based hybrid trust model is proposed and compared with a multi-facet-based trust model. The result suggests that it always selects a low-risk action which is different from what the trust-based approach suggests. However, this work is only suitable for a clustered architecture which is unrealistic for VANET. Ref [26] uses a Bayesian inference-based direct trust and recommendation trust to calculate the final trust. This model finds the confidence of direct trust to avoid the costly recommendation trust calculation. Though this approach is compared with two models, they consider only packet drop and interception as malicious behaviours.

Many blockchain-based trust models are also present in the literature. Here, we only highlight two approaches though our proposed model is not blockchain-based. Ref. [27] presents a three-layer blockchain-based trust model for VANETs using Dirichlet distribution, regression, and revocation. They consider simple, slander, and strategic attacks along with both normal, and malicious servers. However, the work does not reward benevolent activities from vehicles. Also, in [28], the authors present a decentralized blockchain-based trust model which selects a message evaluator through RSU collaboration. The approach finds the rating for messages, the sender, and the evaluator. Next, they calculate the global trust of a node based on the rating, and message quality. They preserve trust data in the blockchain and use a consensus process to insert blocks. They claim that their approach can prevent Sybil, message spoofing, bad-mouthing, and ballot-stuffing attacks. However, this model is not compared against other trust-based models.

## 3. PROPOSED SYSTEM

Ideally, a trust model promoting security should incur little or no extra burden in terms of computational and communication cost. In a VANET, vehicles typically meet each other randomly and fleetingly. Thus, there is little time for decision-making based on trust. With a receiver-side trust model, vehicles with a poor trust score can still send messages although these will typically be ignored by receivers once their trust level is verified. However, this takes time, so vehicles may unnecessarily experience events such as traffic jams. To this end, our research proposes a novel sender-side trust management framework that reduces the amount of trust information passed over a VANET and blocks untrusted transmission attempts. It uses a sender-side TPD on vehicles to prevent unauthorized access and regulate transmissions based on the level of trust. Different classes of messages are created that associate different trust thresholds to permit their broadcast. Drivers improve their trust score from valid announcements, forwarding, beaconing, and clarifying events to an RSU. Also, a driver builds trust from RSU rewards whenever "wins a dispute" over another vehicle. The beaconing reward is only given when a driver is not blocked, and trust is less than 0.5. The TPD also punishes drivers when announcements are delayed, or the vehicle is travelled more than a threshold distance besides the arrangement of RSU punishments whenever a driver "loses a dispute" to another vehicle. Since announcements are regulated by the sender's TPD, receivers can believe messages and the sender's trust instantly.

## 3.1. System Assumptions

The framework assumes that the security of the TPD is beyond the scope of the proposed framework as it relates to physical layer protection. Also, we do not consider other security aspects with this trust framework as we believe that existing security techniques can address authentication, privacy, and integrity. A security approach that supports these functionalities could be incorporated with the framework to confirm the authenticity of the driver and/or messages with other entities, secure the privacy of the driver, and any Hash Message Authentication Code (HMAC) for achieving integrity. For example, when a driver registers with the TA, the driver can obtain private and public keys to encrypt and decrypt messages and can obtain a pseudo-identity associated with his driver ID for securing privacy with other drivers, and HMAC can be used to maintain integrity [1]. RSUs, official vehicles, and the Trust Authority (TA) are also considered fully trustworthy. Both the TA and TPDs are governed and owned by the Road Transport Authority (RTA). The resilience of the TA infrastructure is beyond the focus of this work. We assume a driver has a built-in dashboard with designated touch buttons to display the classes of messages available given his/her current trust score and to generate specific emergency events for other vehicles. Furthermore, a TPD can access GPS to determine the location of the vehicle.

## 3.2. Registration, Blocklisting, and Redemption

Drivers may register themselves with the TA directly using an online form with the vehicle plate number as vehicle ID and driver's license number as driver ID. Since this is an external means, this is outside the scope of the framework. Alternatively, if the system chooses to send a registration message from the driver when they start initially; then the RSU forwards the registration and confirmation of registration to and from the TA.

RSUs send the decisions of disputed events to the TA to store in a driver profile database which keeps driver and vehicle information, the event information, and the reward/punishment. When the TA receives a decision on a disputed event, then it searches the driver profile database. If three malicious events have been found in a limited timeframe, the TA sends a blocking confirmation message to the RSUs in the driver's vicinity. When the vehicle receives a blocking confirmation message, the TPD blocks further access of the driver and acknowledges the blocking confirmation message with the TA via an RSU. The blocked driver can only send/receive beacons into the VANET. Additionally, a blocking message can be generated from the vehicle's TPD when the driver's trust score crosses the lowest acceptable trust limit. This message is forwarded by an RSU to the TA and the same mechanism is followed to block the access of traffic events for the driver in the VANET. By default, based on experimentation, regular drivers obtain access to traffic events in the VANET with a trust score between 0.06 to 0.9. Whenever the trust score becomes lower than 0.06, an external mechanism requires the driver to communicate with the TA to obtain redemption from blocking. We assume this is within the jurisdiction of the road transport authority (RTA) and may involve issuing a monetary penalty or other sanctions.

### 3.3. Framework Components

Both regular vehicles and official vehicles are present within the framework. The framework is extensible so that more vehicle classes could be added. The approach supports the same vehicle accommodating multiple drivers via individual driver trust management. The actions and responsibilities of each vehicle type are limited to their role. Every vehicle is pre-equipped with a built-in On-Board Unit (OBU), comprising a Global Positioning System (GPS) for location access, a transceiver to communicate with other entities, and a TPD that manages the trust and regulates transmissions. We define the following actors based on their roles:

- Senders: are drivers that can originate both true and untrue announcements relating to an incident, such as an accident, subject to their trust score. For a true announcement, the trust manager within the TPD rewards the driver if the claim is not disputed within a given time.
- Reporters: are drivers that refute an announcement of a sender and receive a reward or punishment if the challenge is confirmed or dismissed, respectively. If they do decide to make a report, they may do so either truthfully or falsely. Failure to make a report carries no penalty.
- Receivers: are drivers that receive messages from any entity and relay them automatically provided the hop limit is not reached and their trust is sufficient.
- Clarifiers: If a dispute is detected at an RSU, the RSU transmits a query message seeking clarification concerning a disputed incident. Vehicles that receive this message can choose to answer the query, i.e., to respond to the RSU, confirming or denying that the incident has taken place, or ignore it. If they respond, they are considered clarifiers.
- Road-Side Units: are automated units that receive information from senders, reporters, and clarifiers, either directly or via intermediate vehicles that rebroadcast the received messages. If information from multiple senders, reporters, or a combination of these conflicts, then the RSU will rule on the dispute. RSUs act as an intermediary between the vehicles and the TA.
- Trust Authority: is the ultimate authority in this framework which validates registration and blocklisting of drivers. The TA blocklists a driver whenever it receives a blocklist message initiated from a TPD or if it finds the three malicious events (3ME) for the same driver within a configurable timeframe. The TA then replies with a blocking confirmation to the RSUs in the vicinity of the last disputed event to reach the vehicle's TPD. Incidents reported by RSUs are saved by the TA in an incident database including the location, timestamp, and incident information. The TA also maintains a driver profile database containing the reward/punishment history of drivers.
- Official Vehicles: This framework considers police, ambulance, and fire service vehicles as official vehicles. Their primary task is to respond to emergency issues on roads by cooperating with RSUs. They are always trusted.

### 3.4. Trust Evaluation Mechanism

This framework sets an initial trust score of 0.45 for regular vehicles to avoid the cold start problem and with the expectation that they will achieve a trust score of 0.5 relatively quickly so that they can then announce all events that the framework supports. Regular vehicles are considered trusted when their trust score is 0.5 or higher. A regular vehicle's trust score cannot go above 0.9 or below 0.05. The trust $T$ of a regular vehicle $i$ is expressed by Eqn (1).

$$T_i = \{t \mid t \in R \mid 0.05 < t \leq 0.9\} \qquad (1)$$

The following rules govern the actions of regular vehicles considering T as trust. If $T \leq 0.05$ (blocked state) of a driver and/or vehicle, then TPD sends a blocking message automatically to the TA, and the vehicle can only generate periodic beacons. Then the TPD waits and blocks network access for all traffic events upon blocking confirmation from the TA. If $0.05 < T \leq 0.25$ (not trusted state), then the vehicle can announce periodic beacons as well but cannot forward events from others. If $0.25 < T < 0.5$ (lowly trusted state), the vehicle can send beacons, make limited announcements, and can forward events from others. If $0.5 \leq T \leq 0.9$ (trusted state and highly trusted when (T = 0.9)), the vehicle can forward and announce all classes of events. If regular vehicles spread untrue messages multiple times, then they receive incremental punishments from RSU. If this count becomes three for severe situations like false accident announcements, then the network access for the driver is blocked. Since the framework considers official vehicles, they are assigned a higher trust score over regular vehicles (i.e. T = 1.0) as regular vehicles should not be trusted more than an official vehicles.

We envision a driver's dashboard as consisting of a set of buttons for supported actions in the framework. Appropriate buttons can be pressed relevant to a specific type of road incident. There are three main classes of messages in the hierarchy. The lowest class consists of beacons and "wave" service announcements, though a blocked driver cannot use the "wave" service facility. The next class of messages consists of announcements of poor road conditions, debris, road defects, and so forth. These can only be broadcasted by drivers with a trust score greater than 0.25. The highest class of messages consists of announcements for accidents, traffic jams, road closures, etc., as well as untrue attack reporting messages. To announce a message from this class, a driver needs to have a trust score of at least 0.5. Algorithm 1 shows the announcement, retransmission, relaying, feedback, and reporting activities of regular vehicles. In Algorithm 2, the TPD trust update and blocking management are shown for regular vehicles. Notations and symbols for Algorithms 1 and 2 are provided in Table 1.

Table 1. List of Notations

| Notation | Meaning | Notation | Meaning |
| --- | --- | --- | --- |
| $RSU_r$ | $r^{th}$ RSU | HL and $RT_L$ | hop limit and retransmission limit |
| $T_s(D_s(V_s))$ | $trust_s$ of driver s of vehicles | $Reward_f$, $Reward_{clar}$, and $Reward_{unt\text{-}atck}$ | reward for forwarding, clarification, and reporting |
| $V_{rep}$, $V_{rec}$, & $V_{trust\text{-}cla}$ | reporter, receiver, and trusted clarifier vehicle | $LowTrust_{msg}$ | forwarding is not possible with low trust |
| $timer_{reward\text{-}withhold}$ | when to process reward/punishment | $timer_{bilst}$ | to check blocking condition |
| $evt_e$ and $untrue(evt_e)$ | traffic event and reporting the $evt_e$ | driver_List | registered driver list |
| $RSU_{clarif\_query}$ | clarification query from RSU | $Trust_s$, $Trust_d$ | saved and initial trust |
| $T_{dis}$, and $T_{int}$ | time threshold to send feedback and to report $evt_e$ | Complaint_List | list of reported announcements |
| TTL, and $M_{cls}$ | Time-To-Live, class of messages | longDelayed | driver delayed than the upper limit |
| $ATT(M_{cls})$ | associated trust threshold of $M_{cls}$ | $Msg_{block}$ & $Msg_{block\text{-}conf}$ | blocking and blocking confirmation message |
| $Rew_r/Pun_r$ and $Rew_{tpd}/Pun_{tpd}$ | reward/punishment from a $RSU_r$, and TPD | PosDiff | distance between the event and announcement location |

---

**Algorithm 1. regular vehicle traffic event management**

**Input**: Driver ID, Vehicle ID, events, trust of drivers, hop and retransmit limit
**Output**: controlled broadcasting, relaying, reporting, and sending feedback

1. **case** *eventType* of
2. *witnessed-event:*// to warn others.
3.   **if** $(T_s(D_s(V_s)) \geq ATTL(evt_e))$
4.     $D_s(V_s)$ prepares and broadcasts the $evt_e$
5.     Send metrics to TPD to find $Rew_{tpd}/Pun_{tpd}$
6.   **end if**
7. *reported-event:*// to report the received event.
8.   **if** ($V_s$ decides $evt_e$=false) and ($V_s$ visits event place within $T_{int}$) and $(T_s(D_s(V_s)) \geq 0.5)$
9.     Send $untrue(evt_e)$ towards RSU
10.     Notify TPD to add $T_s = T_s + Reward_{unt\text{-}atck}$
11.   **end if**
12. *relayed-event:*// to relay event up to hop limit.
13.   **if** ($V_s$ gets an $evt_e$ or an $untrue(evt_e)$ from a $V_{rep}$ first time)
14.     **if** ($V_s$ sends $evt_e$ or $untrue(evt_e)$)
15.       Return
16.     **end if**
17.     **if** $(T_s \in (T_s > 0.05$ and $T_s <= 0.25))$
18.       Send a $LowTrust_{msg}$
19.     **else**
20.       **if** TTL($evt_e$ or $untrue(evt_e) \geq$ HL)
21.         Stop resending $evt_e$ or $untrue(evt_e)$
22.       **else**
23.         Resend $evt_e$ or $untrue(evt_e)$ up to HL
24.         Notify TPD to add $T_s = T_s + Reward_f$
25.       **end if**
26.     **end if**
27.   **end if**
28. *retransmit-event:* // to repeat the broadcasting
29.   **if** (no_of_time $\leq RT_L$)
30.     Resend $evt_e$
31.   **end if**
32. *feedback-event:* // to send feedback.
33.   **if** ($V_s$ receives a RSU query about $evt_e$)
34.     **if** ($V_s$ is a $V_{rep}$ or is the sender of $evt_e$)
35.       Return
36.     **end if**
37.     **if** TTL($RSU_{clarif\_query}$) < HL
38.       Resend $RSU_{clarif\_query}$ message
39.     **end if**
40.     **if** ($V_s$ visits the event location within $T_{dis}$)
41.       Send feedback

42.      Notify TPD to add $T_s = T_s +$ reward$_{clar}$
43.     **end if**
44.    **end if**
45.    **end case**

---

Algorithm 2: trust and blocking management at the TPD

**Input:** Announced $evt_e$, reporting status of $evt_e$, Msg$_{block-conf}$, PosDiff, delay, Rew$_r$/Pun$_r$,
**Output:** Trust update and access blocking.

1. **case** eventType **of**
2. *periodic-blocking-checker*://
3. **if** $D_s(V_s)$ is unblocked) and (timer$_{blist}$ expires) and ($T_s(D_s(V_s)) <=0.05$))
4.    TPD$_s$ issue a Msg$_{block}$ to reach TA.
5. **end if**
6. **if** Msg$_{block-conf}$ comes from TA for $D_s(V_s)$
7.    Disable the network access for $D_s$
8. **end if**
9. *RSU reward/punishment:* // add with trust.
10. **if** (Rew$_r$/Pun$_r$ from an RSU$_r$ for the $D(_s(V_s))$)
11.    $(T_s(V_s(D_s)))= (T_s(V_s(D_s))+Rew_r/Pun_r$
12. **end if**
13. *metrics*: // Applies reward withholding.
14. **if** ($evt_e$=false by receiving a complaint)
15.    reward$_{evt-e}$ =0
16. **else**
17.    **case** [PosDiff | D]
18.    0<PosDiff<300m | 0<D<15s:
19.       reward$_{tpd}$ =0.08.
20.    301<PosDiff<500m | 16<D≤30s:
21.       reward$_{tpd}$ =0.06.
22.    501<PosDiff <800m | 31<D≤60s:
23.       reward$_{tpd}$ =0.05.
24.    801<PosDiff<1200m | 61<D≤120s:
25.       reward$_{tpd}$ =0.01.
26.    1201<PosDiff<1500m | 121<D ≤150s:
27.       reward$_{tpd}$ =-0.01.
28.       longDelayed=true
29.    PosDiff >1500m | D>150s:
30.       reward$_{tpd}$ =-0.05.
31.       longDelayed=true
32.    **end case**
33.    **if** longDelayed=true
34.       call reward/punishment process immediately
35.    **else**
36.       start timer$_{reward-withhold}$ to process reward
37.    **end if**
38. **end if**
39. *TPD reward/punishment*://add with trust
40. **if** (broadcasted msg_id ∈ Complaint_List)
41.    Update $T_s(D_s(V_s)) = T_s(D_s(V_s))$
42.    Return
43. **else**
44.    Update $T_s(D_s(V_s)) = T_s(D_s(V_s))$+reward$_{tpd}$
45.    **if** ($T_s(D_s(V_s))$)>0.9
46.       Update $T_s(D_s(V_s))$=0.9
47.    **end if**
48.    **if** ($T_s(D_s(V_s))$)≤ 0.05
49.       Update $T_s(D_s(V_s))$=0.05
50.       Start timer$_{blist}$
51.    **end if**
52. **end if**
53. *complaint-on-broadcasted-event*:// save report.
54. **if** the broadcasted msg_id has a complaint
55.    Save the complaint into the Complaint_List
56. **end if**
57. *driver-change-event*:// to change driver
58. Extract the driver_name $D_s$
59. **if** ($D_s$ exists in the driver_List)
60.    Use $D_s$ as the current driver and Trust$_s$
61. **else**
62.    Add $D_s$ in driver_List and $T_s$=Trust$_d$
63. **end if**
64. **end cas**

---

A potential reward is initially assessed at a TPD and then withheld for a period before adding it to the current trust. During this time, a TPD checks whether any complaint has been raised by any reporter. Rewards are calculated based on message accuracy (no complaint), location difference, and delay/responsiveness. Thus, the framework promotes emergency event announcements at the earliest possible opportunity. The distance a vehicle moves between the event location and the vehicle's current position is passed to the TPD to determine the location difference. Delay is calculated as the difference between the announcement and the observation time on the road. The TPD uses this information to assess the reward/punishment for the announcement. Also, the framework suggests a vehicle should not travel more than 500 meters or the next traffic signal to earn a higher reward from the announcement. The trust $T_i$ is updated inside the TPD using Eqn (2).

$$T_i = T_{i-1} + R_i \text{ "or" } P_i \qquad (2)$$

Here, Ti is the revised trust score of a vehicle after adding a reward/punishment to its current trust Ti-1. Ri "or" Pi is the estimated reward or punishment for the $i^{th}$ message announcement. The set of rules used by the TPD for deciding the appropriate reward/punishment magnitude for a given announcement is shown in the "metrics" event handling in Algorithm 2. In the reward/punishment assessment, the accuracy of the announcement, delay, and location difference is considered and then the current trust is updated using the assessed reward/punishment. The assessed reward is then withheld for a period and the reporting status of the announced message is checked before the trusted update. If the message is accurate but the driver exceeds the thresholds, then no reward or a nominal punishment is issued. But the assessed punishment is deducted immediately from the current trust. However, if the message is inaccurate (i.e., a complaint received during the reward withholds period), then the driver receives no reward for the announcement from the TPD and defers the reward/punishment decision to the RSU. As a VANET is a time-critical system, vehicles should disseminate information promptly. Thus, the delay and distance traveled are given prime importance in the reward calculation besides the message accuracy.

### 3.5. Functional Diagram of the Proposed Framework

In Figure 2, assume a driver sees an incident and wishes to announce it. The framework first checks the trust score from the TPD and determines if the action is eligible with the driver's current trust. If this test is passed, then the driver announces the incident. Other "receivers" forward it up to the configurable hop limit. As this is an original announcement, the driver is classified as a "sender". If this announcement reaches subsequent drivers who visit the same location later, they can notice whether the said event occurs or not. However, if any driver believes the announcement to be untrue, that driver can send a complaint to the RSU. When the RSU receives this complaint, the RSU requests "trusted clarifiers" to respond, confirming or denying the claim.

It collects feedback from these trusted clarifiers who have recently visited the event location. After this, the RSU rules on the validity of the event and penalizes or rewards the respective vehicles. An RSU always informs the TA of the outcome of a dispute, which could be a driver being malicious. It is then up to the TA to check prior behaviour for three malicious activities from a specific driver over a configurable time and block this driver. The TA sends a blocking confirmation message to RSUs in the vicinity from where the TA receives the last dispute decision. These RSUs forward it to the concerned vehicle which receives and acknowledges the instruction. Alternatively, if a driver's trust reduces to 0.05, the TPD generates a blocking request message which a nearby RSU forwards to the TA. Then the TA blocks this driver and informs the respective TPD via the RSUs in the vicinity of the vehicle.

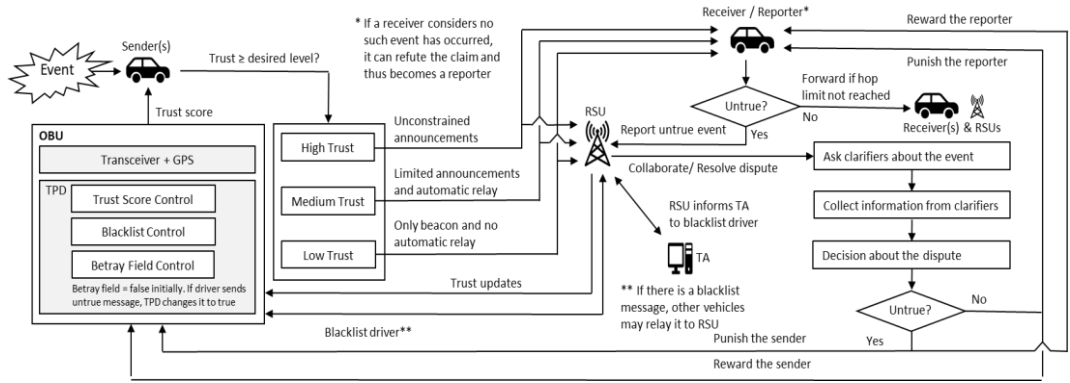

Figure 2. Functional Diagram of Proposed Framework

### 3.6. RSU Traffic Event Management/Functionality

RSUs always listen to traffic events and share them as necessary based on their severity. RSUs receive regular beacons from vehicles. In response, RSUs send beacons periodically to notify of their roadside existence so that other entities can request services. When RSUs receive an emergency traffic event message from a sender vehicle, they rebroadcast the same towards neighboring vehicles so that oncoming vehicles whose route includes the problematic road may avoid it. RSUs also share certain events with nearby RSUs so that vehicles in a greater region may avoid the problematic road, if appropriate. For some events, RSU will continue to periodically announce it until they receive notification from an official vehicle to confirm the event is resolved. When a traffic event occurs and if the RSU receives an AttendingBY-$V_{off}$ message from an official vehicle, the RSU can confirm the event has occurred. The RSU continues to announce the traffic event periodically until the reception of a sorted traffic event from the official vehicle. When the event sorted message arrives, the RSU stops rebroadcasting the original traffic event towards oncoming vehicles. Rather it starts broadcasting only the sorted traffic event up to a retransmission limit as well as forwarding this message to nearby RSUs based on the severity of the original traffic event. RSUs rebroadcast and forward traffic incidents to the TA from sender vehicles besides storing traffic incident information until it is resolved. Each local service point, for example, petrol pumps, and parking is registered in advance with the nearby RSU. Whenever any vehicle sends any query seeking information regarding any service, the local RSU sends a reply to the service query containing the information of queried service or it says it has no information if it does not know.

An RSU assigns a fixed amount of reward and sets the punishment for disputed announcements using an Incremental Punishment Policy (IPP). An RSU forwards the decision of a disputed event to the TA. Then TA checks the malicious event count for relevant drivers. If the 3ME condition holds then the TA sends a blocking confirmation message to the RSUs in the vicinity of the last disputed event. After that, these RSUs broadcast this message to the vehicle. Besides these functionalities, an RSU also resolves disputes when untrue attack messages arise which is illustrated in Section 3.7.

### 3.7. RSU Untrue Message Detection

If an RSU receives conflicting information from a sender and a reporter, it initiates a "collaboration" process to determine the validity of the disputed event. To this end, first, an RSU broadcasts a send-a-reply message to all trusted clarifiers in the vicinity including possible

International Journal of Security, Privacy and Trust Management (IJSPTM) Vol 12, No 1, February 2023

official vehicles and waits for a timer to expire when the feedback collection is finished, as depicted in Algorithm 3. Notations and symbols for this algorithm are listed in Table 2.

Table 2. List of Notations.

| Notation | Meaning |
| --- | --- |
| $evt_e$-sorted | sorted event $evt_e$ |
| untrueHandledList | list of disputed cases which $RSU_r$ has decision |
| send-a-reply($evt_e$) | ask $V_{cla}$s to send feedback on $evt_e$ |
| $Loc_{req}(service_i)$ and $Loc_{rep}(service_i)$ | location of $i^{th}$ service query and $i^{th}$ service reply |
| $RSU_{r\{known\ service\}}$ | known services at $RSU_r$ |
| $timer_{sc}$ | timer to collect feedbacks by an RSU |
| $reply_{off}$ | reply from $V_{off}$ |
| untrue_id | untrue_attack message id |
| AttendingBY-$V_{off}$($evt_e$) | attending $evt_e$ by a $V_{off}$ |
| rewV, punV | rewarded and punished vehicle |

**Algorithm 3: RSU untrue attack handler**

**Input:** untrue attack, feedback message, trust, lists to save events
**Output:** initiate feedback collection for a timer, find rewarded/punished vehicle

1. **while** running
2.   **case eventType** of
3.   *untrue attack:* // deals with untrue attacks.
4.   **if** unique *untrue*($evt_e$) from $RSU_s$/$V_s$
5.     Insert into untrueAddedList
6.     **if** *untrue*($evt_e$) ∈ untrueHandledList
7.       Return
8.     **else**
9.       Insert into untrueHandledList
10.     **if** *untrue*($evt_e$) from a $V_s$
11.       Broadcast a *send-a-reply($evt_e$)*
12.       Start a $timer_{sc}$ to collect feedbacks
13.     **end if**
14.   **end if**
15.   **else**
16.     $RSU_r$ receives an *untrue*($evt_e$) from a $V_{off}$
17.     Call rew-pun-generator($V_{off}$, $V_s$)
18.   **end if**
19.   *feedback:* // collect all the feedbacks.
20.   **while** ($timer_{sc}$ is not expired)
21.     **if** unique feedback $f_u$ from $V_{cla}$ is for $RSU_r$
22.       Insert in vector <$f_0, f_1, \dots f_n$>
23.       **if** $f_u$ is from a $V_{off}$
24.         **if** $f_u$ is the same as the $V_s$'s event
25.           Call rew-pun-gen($V_s$, $V_{rep}$)
26.         **else**
27.           Call rew-pun-gen ($V_s$, $V_{rep}$)
28.         **end if**
29.         Update rewardList and punishmentList
30.       forwardMsgtoRSU$_s$(decision_untrue)
31.       **if** count(3ME($V_s$ or $V_{rep}$)) ≥ 3
32.         Send a $Msg_{block}$(count(3ME($V_s$ or $V_{rep}$) ≥ 3)) to TA
33.       **end if**
34.     **end if**
35.     **else**
36.       The feedback is for different $RSU_s$
37.     **end if**
38.   **end while**
39.   *decision-of-untrue:* // to resolve dispute
40.   **if** $timer_{sc}$ expires
41.     **if** the *untrue*($evt_e$) has a decision
42.       Return
43.     **else**
44.       Sum=0
45.       **case feedbackType** of
46.         Positive:   $F_i = 1$
47.         Negative:  $F_i = -1$
48.         Unsure:    $F_i = 0$
49.       **end case**
50.       **for** each $F_i$ from feedback vector <$F_n, T_n$>
51.         Sum += $T_i * F_i$
52.       **end for**
53.       **if** Sum>0
54.         $V_s$ send true event, $V_{rep}$ send false report
55.         Call rew-pun-generator($V_s$, $V_{rep}$)

| | | | |
|---|---|---|---|
| 56. | **else if** (sum<0) | 64. | **if** (sum>0 or sum<0) |
| 57. | $V_s$ send false event, $V_{rep}$ send true report | 65. | forwardMsgtoRSU$_s$(*untrue_dec*) |
| 58. | Call rew-pun-gen ($V_{rep}$, $V_s$) | 66. | **end if** |
| 59. | **else** | 67. | Clear the vector<feedback> on *untrue_id* |
| 60. | Undecided conflict | 68. | **end if** |
| 61. | Insert attack into unresolvedUntrueList | 69. | **end if** |
| 62. | Send an unresolvedUntrue(*evt$_e$*) to a $V_{off}$ | 70. | **end case** |
| 63. | **end if** | 71. | **end while** |

Sender(s) and reporter(s) involved in the dispute are not permitted to participate in this clarification process. It is reasonable to consider that there are some trusted vehicles around the event. Also, there may be several malicious vehicles, as considered in [2, 13]. The effect of malicious feedback will be nullified when the true feedback outweighs the malicious feedback while taking a decision. In this framework, feedback can only be generated by trusted clarifiers with trust scores greater than 0.5 and official vehicles. The possible feedback messages are 'YES' or 'NO'. Eligible vehicles that respond, known as clarifiers, reply 'YES' if they had visited the event location recently and confirmed the event or 'NO' if they had visited the event location and did not see the event. In some cases, drivers neither notice the event nor visit the event location in the recent past. These drivers will simply ignore the RSU query. Also, official vehicle feedback is treated as the decider for a dispute which bypasses the collaboration process since collected feedbacks from the trusted clarifiers are not used in forming a decision. When an RSU receives official vehicle feedback in Algorithm 3, it instantly invokes the reward-punishment generator as shown in Algorithm 4.

| Algorithm 4: rew-pun-gen (rewV, punV) | |
|---|---|
| **Input**: rewarded vehicle, punished vehicle | |
| **Output**: send reward/punishment message and blocking message to TA, if required | |
| 1. **while** running | 4.  **if** (rewV!=$V_{off}$) |
| 2.  *reward/punishment:* //estimate reward or punishment for disputed event | 5.   Send the reward_msg(rewV) |
| | 6.  **end if** |
| 3.  Store reward(rewV) and punishment(punV) in rewardList and punishmentList | 7.  Send the punishment_msg (punV) |
| | 8.  Call forwardMsgtoTA(untrue_dec) |
| | 9. **end while** |

The RSU dispute resolution mechanism in Algorithm 4 uses these feedbacks to decide the truthfulness of a dispute. Here, the RSU performs a sum of product calculation of the feedback and trust of the clarifiers to decide on the disputed event. For example, suppose a vector of feedback is ('YES', 'YES, 'NO', 'NO', 'YES') which are represented programmatically as (1, 1, -1, -1, 1) and the clarifier's corresponding trust scores are: (0.5, 0.7, 0.65, 0.68, 0.9), then the RSU decides by using Eqn (3). It should be noted that only trusted clarifiers can join the collaboration process. Generally, Eqn (3) can be expressed as in Eqn (4) for n feedbacks collected from n trusted clarifiers, where $F_i$ is the $i^{th}$ feedback and $T_i$ is the $i^{th}$ clarifier's trust score.

$$Decision = [1*0.5] + [1*0.7] + [-1*0.65] + [-1*0.68] + [1*0.9] \quad (3)$$
$$Decision = \sum_{i=1}^{n} F_i * T_i \quad (4)$$

If the outcome is positive, then the RSU decides the sender has disseminated a true event and thus receives an RSU reward; the conflicting reporter(s) receive an RSU punishment. If the outcome is negative, the converse actions are followed. When the decision is reached, the RSU calls the reward-punishment generator, shown in Algorithm 4. During the punishment assessment, an incremental punishment policy (IPP) is applied to influence the future good behaviour of drivers. However, for an unresolved issue when RSU has no feedback data or Decision=0 in Eqn. (4), the RSU stores them in an unresolved dispute list and later may ask an official vehicle to inspect the event location physically and report its findings so that the RSU can take an action on the dispute. It should be noted that if during the collaboration process, any official vehicle receives an RSU message, but they have not visited the disputed event location recently, then they reply with a far-from-event message. However, if the RSU receives a decisive message from an official vehicle, then it always decides on the event using this message and bypasses the collaboration mechanism.

## 4. IMPLEMENTATION

The framework is implemented in Veins 5.0 [29] which is a tightly coupled framework for simulating VANETs comprising the SUMO traffic simulator [30] and OMNeT++ discrete event network simulator [31]. We extend Veins in several respects. First, we create four types of vehicles, namely: "official" vehicles (police, ambulance, and fire service) and regular vehicles. A TA module has been created that registers and blocks drivers. Additionally, the TA unit keeps a driver's most recent reward/punishment history in a driver profile database to facilitate the blocking of malicious drivers and to record incident information (location, timestamp, incident) in an incident database. In addition to this, an RSU internetwork is developed which also connects to the TA unit via wired communication. Inside each RSU, besides event management, a dispute resolution process is implemented to detect untrue/inconsistent attacks.

### 4.1. System Model and Environment

As there are four types of vehicles, we have created four distinct modules in OMNeT++ and the C++ implementation of them according to the functionality specified in Section 3. A TPD module is added to regular vehicles which primarily implements the trust update and access blocking of each driver. The module can exchange messages with the vehicle application layer using an internal connection. When drivers broadcast messages, they send the delay, and the location of the event from the current location of the vehicle identified on the map. Also, the TPD can check the current location of the vehicle and then determine the amount of reward/punishment from an announcement using Algorithm 2. The TPD withholds rewards for a given period to allow disputes to arise via reporters. Also, the TPD disables the transmission of a blocked driver to stop the generation of event announcements; however, whilst in the blocking state a driver still broadcasts beacons. The vehicle application first reads the trust from the TPD. This trust needs to satisfy the associated trust threshold for the message class to proceed with the announcement. The reward varies for activities like beaconing, forwarding, and broadcasting announcements. Vehicles can only obtain a beaconing reward if they are classified as *not trusted* or *lowly trusted* as defined in Section 3.4. The TPD can also support multiple driver profiles in case different people share a vehicle.

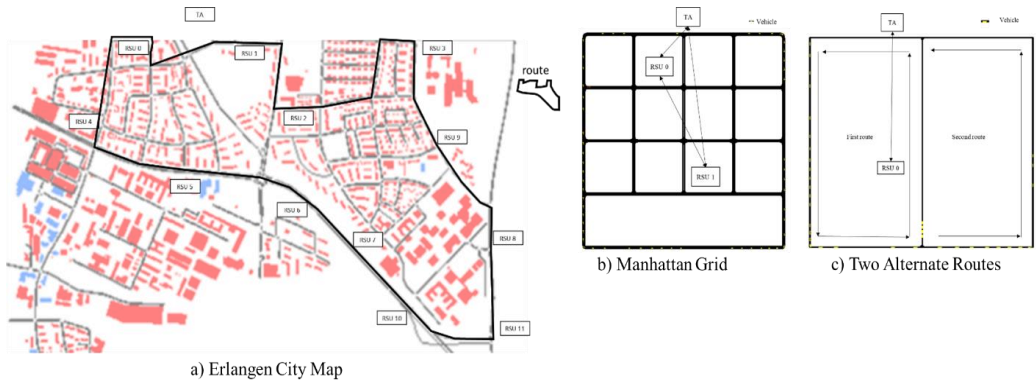

Figure 3. Road Networks Used in the Simulation

The application layer of official vehicles is different than the regular vehicles. Official vehicles respond to RSU queries differently than regular vehicles. Also, when an emergency arises, an RSU gives precedence to their messages over those from regular vehicles. There is a built-in mobility module from Veins that advances all vehicles at regular time steps. The system starts with no vehicles in the terrain model and once a vehicle is added, it remains in the system until the simulation terminates. Once a predetermined number of vehicles enter the system, no more are permitted. The simulation commences by assigning periodic events to specific vehicles and then the resultant data are collected regarding the specific experiment. The framework is simulated using road networks from the Veins default Erlangen city map [29] as in Figure 3a, the Manhattan grid map in Figure 3b, and one alternate route scenario in Figure 3c.

### 4.2. Verification – Thwarting Untrue and Inconsistent Attacks

This framework is verified in the presence of malicious and benign behaviours of trusted vehicles. To this end, this experiment is conducted at least thirty times for 5000 simulation seconds with 10-100 vehicles. One result from these experiments is depicted in Figure 4 to illustrate trust management in the presence of attacks. The horizontal axis represents the simulation time in seconds, and the vertical axis represents the trust score. All vehicles start with a trust score of 0.9. The reward is fixed for a single announcement or RSU interaction which is set to 0.08. and RSU punishments for three untrue announcements are set to 0.1, 0.3, and 0.5 (applying IPP) consecutively. Receivers report an event with a probability of 40% and the event supporting probability P from clarifiers is set to 20%. Vehicle V0 constantly sends messages at 200s intervals (simulation seconds) starting from 100s. Figure 4 records the inconsistent behaviour of V0 with the consistent behaviours of V1 and V2.

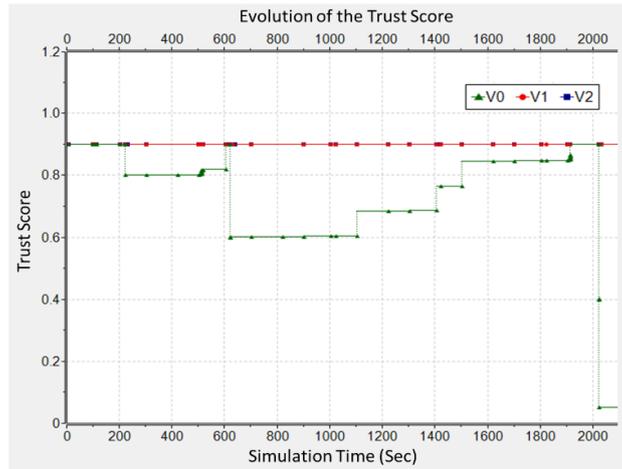

Figure 4. Trust Increment/Decrement with Untrue and Inconsistent Attacks

When V0 sends untrue messages, the RSU punishes it by 0.1 and 0.3 consecutively at 220s and 640s. Conversely, when V0 announces trustworthy messages consecutively at 500s, 1100s 1300s, 1500s, 1900s, the TPD adds a reserve reward at 620s, 1220s, 1420s, 1620s, and 2020s; these rewards are withheld for 120s. V0 receives complaints for message announcements at the 700s, 900s, and 1700s. Thus, the TPD does not add any reward for these announcements. After this, the RSU punishes V0 by 0.5 which is shown by a large reduction in the trust score in the 2020s followed by a blocking message which sets its current trust score to 0.05 irrespective of whatever it previously had. In this way, V0 is blocked from network access by the framework. From this simulation result, it can be concluded that the Incremental Punishment Policy (IPP) will demotivate vehicles from attacking repeatedly. The IPP provides a flexible means of punishing and blocking vehicles for their inconsistent behaviours although they may sometimes announce trustworthy messages in between their malicious activities. As vehicles receive higher punishment in each subsequent untrue announcement, vehicles with inconsistent behaviour will be isolated as well. Additionally, we allow only three malicious actions within the simulation timeframe (for example 5000s) from a trusted driver to become blocked to limit further harmful actions. So, in summary, this trace represents an example confirming the system can successfully detect the inconsistent behaviour of a malicious vehicle and punish it accordingly.

## 5. PERFORMANCE EVALUATION

In this section, the evaluation of the proposed approach is investigated. The simulation model described in Section 4 is used in the presence of varying traffic densities. Clarifiers generate varying percentages of malicious and benevolent feedback to classify events as true negative, true positive, false positive, and false negative observational data. Analysis shows the proposed framework can classify events as expected. This means that when there is more benevolent feedback the RSU can classify an event correctly and vice versa.

This set of experiments considers the generation of varying ratios of positive and negative feedback when classifying a disputed event. Moreover, we show the minor impact of vehicle density on the results as RSUs ignore repeated complaints regarding the same event. As expected, when density increases, more vehicles complain about an untrue event. The RSU forwards the first complaint to nearby RSUs which avoids invoking costly concurrent collaboration procedures at other RSUs. Furthermore, the proposed approach is compared against a reputation approach [8] in terms of response time and communication overhead. The response time is the decision time of receiver vehicles when they receive an event message in the network. A trust management model for a VANET can be considered the most efficient one when receivers can decide about an event in the fastest possible time relative to trust systems where additional computation and communication are required after the arrival of messages. Hence, response time is a good indicator of performance. Another useful metric is the communication overhead since a trust model with lower communication overhead reduces the burden of message transmission and processing. For this reason, a trust model with a lower response time and communication overhead can be regarded as superior to one where these values are higher. Furthermore, these two metrics affect the performance of the network communication, such as channel availability, hence the proposed approach is compared with a reputation approach which suffers from these two factors. Results from the analysis show the proposed framework outperforms the existing one as a receiver vehicle in the proposed framework can decide on the appropriate action without further communication within the VANET. However, the proposed framework requires broadcasting feedback other than the traffic event if a reporter invokes an untrue attack event.

### 5.1. Scenario 1 - Accuracy of the Proposed Framework

#### 5.1.1. Simulation Setup

This series of simulations have been conducted using one predefined route for both regular and official vehicles on the Erlangen city map from Veins [29]. Selected parameters used for conducting the series of simulations are listed in Table 3.

We repeat each experiment five times to collect trial data for every vehicle density and probability of supporting an event. This sample data is then averaged for analysis. We use the probability P to control the support or denial of events from clarifiers through YES/NO responses. For example, with a probability of P=0, clarifiers always send NO. For P=1, clarifiers always send YES. For probabilities of P=0.2…0.8, clarifiers send YES/NO responses accordingly. In this way, the analysis considers varying ratios of benevolent and malicious feedback. We consider, the senders always announce true events in one set of experiments, whereas, in another set, they always announce untrue events. One or more reporter vehicles may send untrue attacks upon reception of these events which are also randomized with a probability of 0.4.

Table 3. Parameters for Simulation Setup.

| Selected Parameter | | Value |
| --- | --- | --- |
| Simulation Parameter | Simulation area | 2.5km X 2.5Km |
| | Number of vehicles | [10, 30, 50, 70, 90, 100] |
| | Number of RSUs | 12 |
| | Speed of vehicles | Max 80 m/s |
| | Simulation time (seconds) | 4000s |
| | Transmission range | 300m |
| | Warm-up period | 700s |
| | Number of event source | 3 |
| | Periodic announcement | At 100s |
| | Event supporting probability | [0, 0.2, 0.4, 0.6, 0.8, 1] |
| | Reward withhold timer | 120s |
| | Collaboration timer | 120s |
| | Initial trust | 0.8 |
| Attacker Model | Untrue and Inconsistent attacks initiated | Applies to regular vehicles |

The subsequent analysis considers P, the probability of being truthful or not, and D, the vehicle density. The possible RSU judgements are given in Table 4.

Table 4. Results Classification Matrix

|  | Predicted true | Predicted false |
|---|---|---|
| A true event | True Negative (TN) | False Positive (FP) |
| A false event | False Negative (FN) | True Positive (TP) |

### 5.1.2. Results and Analysis

In Figure 5, the x-axis represents the vehicle density, the y-axis represents the probability of being truthful or not and the z-axis represents the normalized likelihood of classified cases. We define the normalized likelihood of TN/FP classified cases as the ratio of the average number of classified TN/FP cases to the average number of reported events which the RSU classifies as TN/FP using the dispute resolution process. When the vehicle density increases, the number of reporters also increases. However, throughout the simulations, increasing vehicle density is shown to have only a marginal impact on the results as the RSUs ignore repeated complaints concerning the same event from multiple reporters. This is possible as the RSU which received the first complaint concerning an event forwards notification of it immediately to other RSUs in the vicinity to prevent invoking further costly and redundant collaboration procedures.

Figure 5a shows the TN results for a series of simulations where sender vehicles announce only true events. Overall, as P increases, the possibility of classifying TN cases also increases. This means, the framework correctly classifies disputed events if most clarifiers send truthful feedback. As expected, at P=0, there are no TN cases because all clarifiers deny the original event. Alternatively, at P=1, all reported events are TN as all clarifiers only send YES. The TN cases increase rapidly from P=0.4 to P=0.6 as the proportion of received YES responses is sufficient to support the sender announcement at RSUs. Also, the number of TN cases increases with the rise of vehicle density as the increased traffic supports the sender's announcement. Figure 5b shows the FN results of classifying FN cases where sender vehicles announce only untrue events which show a similar trend like TN cases. Conversely, Figure 5c shows the TP chart for a series of simulations when sender vehicles announce only untrue messages. Overall, as P increases, the possibility of classifying TP cases decreases. That means the approach correctly classifies untrue events if most clarifiers rebuff the original, malicious sender message(s). The rapid fall in TP cases noticed between P=0.4 to 0.6 arises when most clarifiers confirm the untruthfulness of the sender's announcement to the RSU query. Here also the number of TP cases decreases more with increasing vehicle density as the increased traffic leads the RSU to receive more YES responses (maliciously supporting the sender announcement). Finally, Figure 5d shows the false FP when the sender announces only trustworthy messages which also shows a similar trend like TP chart.

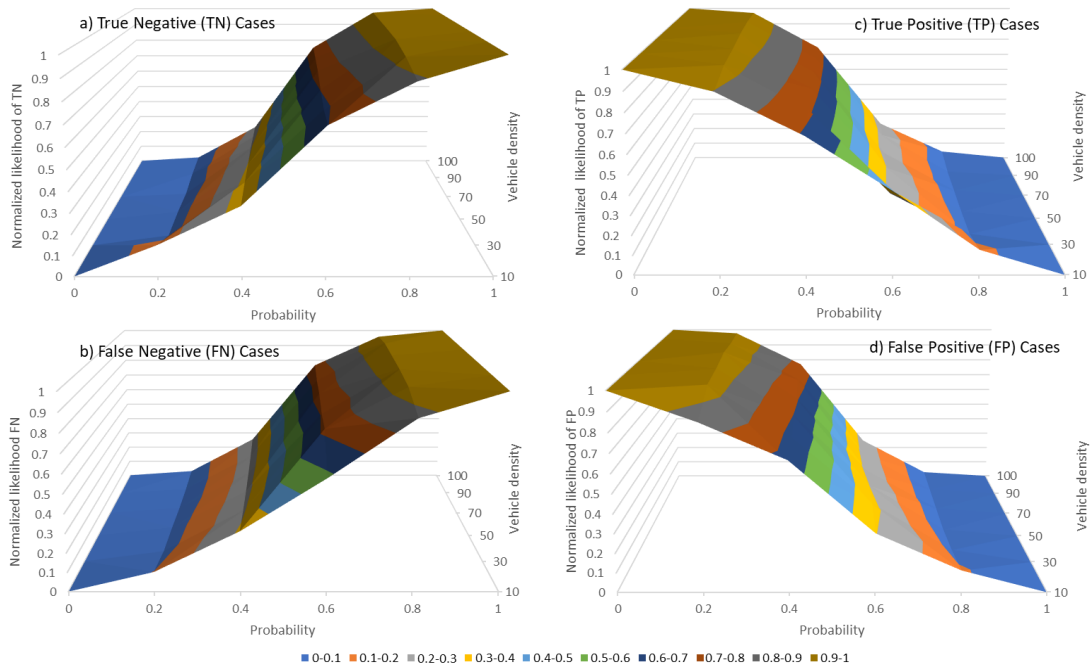

Figure 5. Normalized Likelihood of Classified Cases

Thus, the normalized likelihood of classified TN/FN cases is lower at lower P values and is greater at higher P values than the expected trend. To classify an event as TN/FN, an RSU needs more YES/NO feedback than NO/YES, respectively. The proportion of YES and NO feedback received at an RSU is reflected in the decision which causes the curve to vary nonlinearly with P. Also, the TP/FP curves show a nonlinear relationship for a similar reason.

## 5.2. Scenario 2 – Comparison with Baseline Approach

### 5.2.1. Simulation Setup

We implement a baseline approach [8] alongside our scheme in order to compare message flows. We see that the receiver-side trust evaluation approach suffers badly from communication overhead due to trust metric dissemination as receivers are busy with trust verification after the arrival of messages. In [8] the trustworthiness of a sender is decided using one of the following schemes: majority voting, weighted voting by reputation, and highest reputation level. The feedback is collated at the RSU, and the trust score is subsequently interrogated. This set of experiments runs for 800 simulation seconds and is repeated 10 times to obtain the average number of messages exchanged in the presence of 10 to 70 vehicles. An event is introduced deliberately at 400 seconds in both approaches. In approach [8], all vehicles upon observing the event, announce it. Conversely, in the proposed framework the announcement of an event from one vehicle is adequate with receivers relaying it up to 4 hops.

### 5.2.2. Results and Analysis

In Figure 6, the x-axis represents the number of vehicles present in the simulation, and the y-axis represents the communication overhead for a single event. This framework is compared against the approach in [8] with 30 and 45-second interval timers. It is clear from Figure 6 that the overhead is higher in [8] with both timer durations than in the proposed framework. With a 30s timer in approach [8], the communication overhead is two, three, and four times higher than the proposed framework when the number of vehicles is 50, 60, and 70, respectively. In most situations, the overhead in the proposed framework is significantly lower than the approach [8], which suffers from a higher overhead due to the need of generating feedback towards RSU for regular reputation updates.

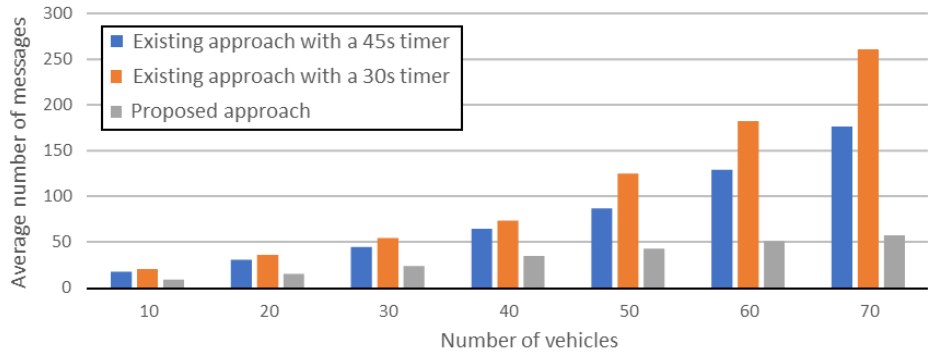

Figure 6. Communication Overhead Comparison

In addition, as expected, the proposed framework is better when we compare its response time against a receiver end-based trust approach. Approach [8] starts a timer for a predefined period i.e., the 30s, to collect additional messages about the same event. When it expires, receivers decide on an event. That is why it suffers from a higher response time. During this time, vehicles may enter a "problematic" road area as they are typically moving fast. On the other hand, the proposed framework quickly decides an event without further communication unless it is disputed. Thus, the proposed framework exhibits a faster response time in comparison with schemes such as [2, 8, 9, 12, 14, 20, 23]**.**

## 6. CONCLUSION

In summary, this research proposes a TPD-based sender-side trust management framework for VANETs. The framework reduces trust metric communications at the expense of equipping every regular vehicle with a TPD. TPD installation is a one-time cost whereas, the circulation of trust metrics is continuous assuming the communication is ongoing. Senders are trusted by default and so response times are reduced at the receivers as trust confirmation is avoided. Additionally, results suggest that the RSUs can successfully resolve any true/false complaints and can detect untrue and inconsistent attacks if the majority of clarifiers send truthful feedback. Furthermore, the framework can be enhanced with additional functionality. For example, a driver profile-based reward/punishment database could be added where historical information concerning driver misdemeanours can tune subsequent rewards and punishments.

**ACKNOWLEDGEMENTS**

This work is supported by the University of Dhaka under the Bangabandhu Overseas Scholarship scheme funded by the People's republic of Bangladesh government for improving quality of education and research.

**AUTHORS**

**Rezvi Shahariar** received his B.Sc. degree in Computer Science from the University of Dhaka, Bangladesh in 2006; and an M.S. degree in Computer Science in 2007 from the same institution. After some time as a lecturer at the University of Asia Pacific, Dhaka, Bangladesh, he is now an Assistant Professor at the Institute of Information Technology, University of Dhaka, whilst pursuing a PhD at Queen Mary University of London. His research interests include wireless network analysis with an emphasis on trust, security in VANETs, and the application of machine learning to security.

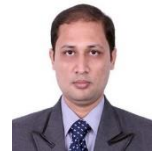

**Chris Phillips** (MIEEE) received a BEng. Degree in Telecoms Engineering from Queen Mary, University of London (QMUL) in 1987 followed by a PhD on concurrent discrete event-driven simulation, also from QMUL. He then worked in industry as a hardware and systems engineer with Bell Northern Research, Siemens Roke Manor Research and Nortel Networks, focusing on broadband network protocols, resource management and resilience. In 2000 he returned to QMUL as a Reader. His research focuses on management mechanisms to enable limited resources to be used effectively in uncertain environments.

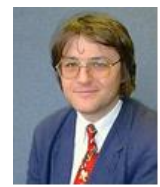